\documentclass{article}

\usepackage{PRIMEarxiv}

\usepackage[utf8]{inputenc} 
\usepackage[T1]{fontenc}    
\usepackage{hyperref}       
\usepackage{url}            
\usepackage{booktabs}       
\usepackage{amsfonts}       
\usepackage{nicefrac}       
\usepackage{microtype}      
\usepackage{lipsum}
\usepackage{fancyhdr}       
\usepackage{graphicx}       
\graphicspath{{media/}}     
\usepackage{orcidlink} 

\usepackage{cleveref}
\title{RedDino: A foundation model for red blood cell analysis}
\author{
  Luca Zedda\,\orcidlink{0009-0001-8488-1612}, 
  Andrea Loddo\,\orcidlink{0000-0002-6571-3816}, 
  Cecilia Di Ruberto\,\orcidlink{0000-0003-4641-0307}, 
  Carsten Marr\,\orcidlink{0000-0003-2154-4552} \\
  \\
  \textit{Department of Mathematics and Computer Science, University of Cagliari, Cagliari, Italy} \\
  \texttt{\{luca.zedda,andrea.loddo,cecilia.dir\}@unica.it} \\
  \\
  \textit{Institute of AI for Health, Helmholtz Munich, Munich, Germany} \\
  \texttt{carsten.marr@helmholtz-muenchen.de}
}

\begin{document}
\maketitle

\begin{abstract}
Red blood cells (RBCs) are fundamental to human health, and precise morphological analysis is critical for diagnosing hematological disorders. Despite the potential of foundation models for medical diagnostics, comprehensive AI solutions for RBC analysis remain limited. 
We introduce RedDino, a self-supervised foundation model specifically designed for RBC image analysis. Leveraging a RBC-tailored version of the DINOv2 self-supervised learning framework, RedDino is trained on an extensive, meticulously curated dataset comprising 1.25 million RBC images from diverse acquisition modalities and sources. Comprehensive evaluations demonstrate that RedDino significantly outperforms existing state-of-the-art models on the RBC shape classification task. Through systematic assessments, including linear probing and nearest neighbor classification, we validate the model's robust feature representation and strong generalization capabilities. Our key contributions are (1) a dedicated foundation model tailored for RBC analysis, (2) detailed ablation studies exploring DINOv2 configurations for RBC modeling, and (3) comprehensive generalization performance evaluation. 
We address key challenges in computational hematology by developing RedDino, a robust and generalizable model that captures nuanced morphological characteristics and represents a substantial advancement in developing reliable diagnostic tools.
The source code and pretrained models for RedDino are available at \href{https://github.com/Snarci/RedDino}{\texttt{https://github.com/Snarci/RedDino}}, and the pretrained models can be downloaded from our Hugging Face collection at \href{https://huggingface.co/collections/Snarcy/reddino-689a13e29241d2e5690202fc}{\texttt{RedDino Huggingface space}}.
\end{abstract}

\section{Introduction}
The hematopoietic process forms the foundation of the blood cell life cycle, where stem cells, through natural aging and mutation, differentiate into various subtypes essential for sustaining bodily functions. Understanding this intricate process is key to unraveling numerous hematological phenomena and diseases~\cite{doulatov_hematopoiesis_2012}. 
In hematology, computer-aided diagnosis has emerged as a tool to tackle critical diagnostic challenges~\cite{sadafi_frontiers_nodate}. Two prominent areas of focus are red blood cell and white blood cell analysis~\cite{chossegros2024improving2,cuevas_improved_2013,deshpande_review_nodate}, both of which rely extensively on imaging-based assessments to derive meaningful insights into a patient’s health.
Blood smear analysis serves as the cornerstone of these investigations, involving the microscopic examination of blood smeared on glass slides. These slides are systematically stained to enhance cellular structures, with staining techniques tailored to specific diagnostic objectives. Variability in staining protocols and imaging acquisition introduces biases~\cite{zedda_deep_2025}, complicating the analysis process and demanding extensive training for medical personnel. Beyond staining, the physical preparation of smears can also slightly alter cell morphology due to the pressure applied during the process. 
For efficient and robust analysis of white blood cells, the emergence of foundation models has significantly advanced the prediction of clinical outcomes~\cite{koch_dinobloom_2024}. These models demonstrate impressive capabilities while addressing critical challenges, such as the batch effect, a common issue in multi-source or multi-patient scenarios~\cite{moor_foundation_2023,zhang_challenges_2024}.
In contrast, red blood cell analysis has yet to fully explore the potential of such advanced technologies. This work seeks to pioneer the development of foundation models tailored for RBC analysis, addressing the nuances and requirements necessary for achieving highly representative and expressive models in this domain. Since training foundation models typically requires massive amounts of data~\cite{oquab_dinov2_2023}, we also contributed to extracting the most complete and representative collection of red blood cell images in different acquisition scenarios.
We demonstrate that, with carefully guided tailoring to state-of-the-art self-supervised learning methods, our family of foundation models achieves state-of-the-art performance across several red blood cell datasets. These models exhibit remarkable generalization capabilities, effectively mitigating the challenges posed by the batch effect.
Our contributions include:
\begin{itemize}
    \item A dedicated family of foundation models optimized for RBC analysis, named RedDino, trained using self-supervised learning.
    \item Rigorous comparative investigation studies to evaluate the effectiveness of DINOv2~\cite{oquab_dinov2_2023} configurations in capturing RBC morphology.
    \item Extensive benchmarking against existing state-of-the-art models, demonstrating superior performance on RBC classification and shape analysis tasks.
\end{itemize}

\section{Methodology}
\subsection{Training Data}

\begin{figure}
    \centering
    \caption{The RedDino training set comprises 56712 original images. We extracted over 3 million single RBC images and more than 1.2 million patches.}
    \includegraphics[width=1.00\linewidth]{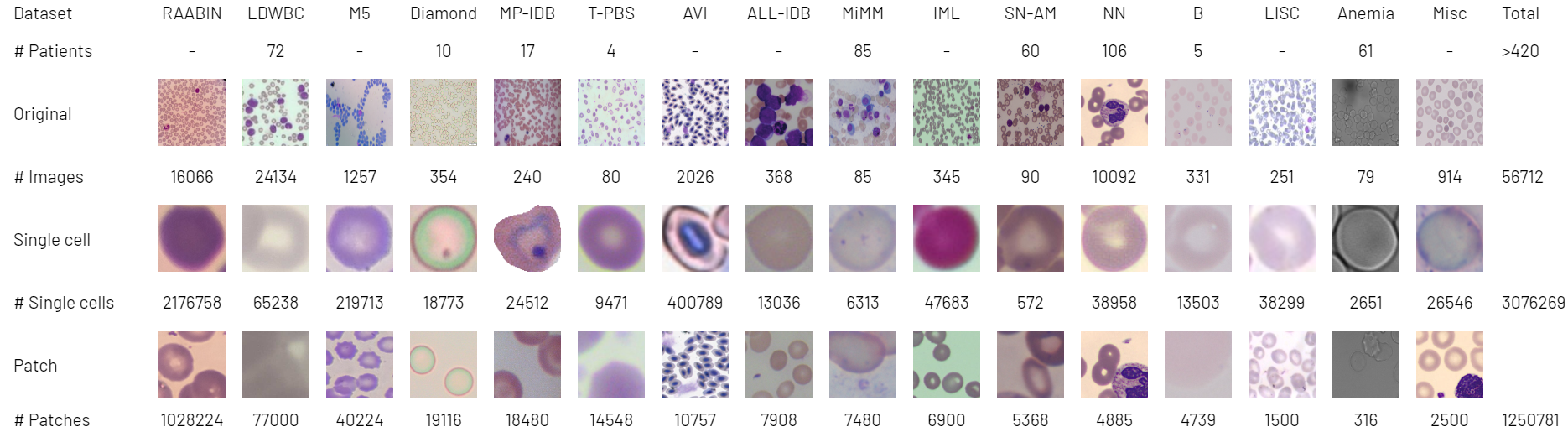}
    \label{fig:crops_train}
\end{figure}

RedDino is trained on the largest collection of publicly available RBC image datasets, spanning various imaging modalities, resolutions, and staining techniques. Leveraging the DINOv2 self-supervised framework, our dataset selection was unconstrained by label availability, a significant advantage given the scarcity of annotated RBC data.
We selected 18 datasets~\cite{DIAMOND,MIMM,SN-AM,IML,chen_accurate_2022,DATASET_B,kouzehkanan_large_2022,ALL-IDB,MP-IDB,MADHLOOM,ANEMIA_DSE,rezatofighi_automatic_2011,BBCD,sidhom_deep_2021,M5,ThalassemiaPBS,AVI}, comprising over 50,000 images from more than 420 individuals (\Cref{fig:crops_train}). To mitigate the natural imbalance between red and white blood cells, we also incorporated datasets containing white blood cell images~\cite{MIMM,MADHLOOM,BBCD}.
To extract training samples, we applied two approaches. The first involved segmentation with a fine-tuned version of CellPose~\cite{pachitariu_cellpose_2022,stringer_cellpose3_2024,stringer_cellpose_2021}, iteratively improved through manual corrections, producing 3,076,269 segmented cells. The second approach involved extracting non-overlapping patches of smear images with a size of 224 by 224 pixels, ensuring aspect ratio preservation and dataset diversity, generating 1,250,781 patches (\Cref{fig:crops_train}). 

\subsection{Testing Data}
To rigorously evaluate model embeddings, we used datasets with diagnostic or morphological labels as out-of-distribution test sets. Specifically, we employed the Elsafty dataset~\cite{ELSAFTY}, the most comprehensive resource for RBC classification, containing 240,000 images across nine classes from four sources, with image distributions of 72517, 52506, 52103, and 63381, respectively. Additional out-of-distribution test sets included the Chula dataset~\cite{CHULA}, comprising 20,875 images across 12 RBC classes, and the DSE dataset~\cite{ANEMIA_DSE}, which consists of 5,659 images spanning eight classes.

\section{Experiments and Results}
\begin{figure}
    \centering
    \caption{RedDino outperforms the baseline models on the weighted F1 score in the linear probing evaluation by removing the Koleo regularizer and applying the Sinkhorn-Knopp algorithm. The evaluation uses source 1 of the Elsafty dataset as the training set and source 2 as the test set.}
    \includegraphics[width=0.8\linewidth]{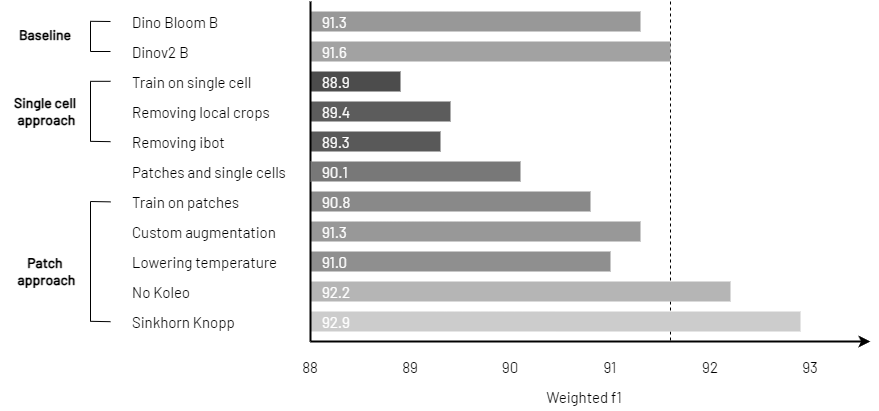}
    \label{fig:ablation}
\end{figure}

\subsection{Development strategies}

An extensive comparative investigation was conducted to evaluate and improve RedDino's effectiveness. Comparisons were made against baseline models, specifically DINOv2 and DinoBloom~\cite{koch_dinobloom_2024}, the latter being the current state-of-the-art feature extractor for hematological data. The experiments were divided into two main approaches: one focusing on training with individual cells and the other on patched smear images. The optimal configuration, which guided the development of all subsequent models, was selected from the best-performing setup.

Since RedDino is designed to serve as a state-of-the-art backbone for various tasks in RBC analysis, its development was guided by evaluating feature quality in RBC shape classification, using the Etsalfy dataset as the primary benchmark. Among the various training strategies explored, the DINOv2 framework was selected due to its strong performance on natural images and robust generalization across downstream tasks.

The evaluation approach involved extracting features from the models and training a linear probing to assess feature quality. Specifically, a logistic regression classifier from the sklearn library was trained on data from source 1 of the Etsalfy dataset and tested on source 2. The results revealed several key findings. First, training RedDino resulted in a performance decline, similar to DinoBloom, when local crops were used in the training process. Second, training on patched smear images instead of individual cells led to significant improvements in model performance. Third, integrating a custom augmentation pipeline, which replaced DINOv2’s pixel-level augmentations with 32-pixel-level augmentations from the Albumentations library~\cite{ALBUMENTATIONS}, further enhanced feature quality.

Two key modifications enabled the model to outperform the state-of-the-art. The first was the removal of the Koleo regularizer, which is critical in natural image scenarios to prevent feature collapse by ensuring uniform feature distribution. However, the regularizer hindered representation quality for RBC images due to the natural uniformity of RBCs in shape and color. Pathological and abnormal RBCs, which should stand out in the feature space, were overly suppressed. The second modification involved replacing the moving average centering with the Sinkhorn-Knopp centering, which improved representation quality (\Cref{fig:ablation}).

The final model configuration, identified as the best-performing approach, is the foundation for developing the full suite of RedDino models. We then proceeded to train a suite of models for 2,000 iterations each, after which performance decreased over time, a pattern not limited to our RedDino models but a well-known phenomenon in foundation model research~\cite{roth_low-resource_2024}.
We finally trained our models using the same hyperparameters as the original DINOv2 while adjusting the batch size to accommodate training on two NVIDIA A100 80GB GPUs. In this setup, the RedDino small model employs a feature dimension of 384 with a batch size of 512 and contains 22 million parameters. The RedDino base model is configured with a feature dimension of 768, a batch size of 384, and 86 million parameters. Lastly, the RedDino large model features a 1024-dimensional feature space, a batch size of 256, and comprises 304 million parameters.

\subsection{Multiple scenario evaluation}
Our downstream tasks aim to evaluate the expressiveness of the features extracted from the RedDino model family while keeping the evaluation scenario as close as possible to the actual use case. 
We chose to focus on the classification task either because ViTs have already demonstrated impressive results in natural image segmentation and detection using adapters~\cite{chen2022vitadapter} or because state-of-the-art hematology models like CellPose~\cite{stringer_cellpose3_2024} and RedTell~\cite{sadafi_frontiers_nodate} have achieved remarkable success.


On the classification side, the heavily imbalanced nature of the data poses a critical challenge. As with most medical imaging problems, many classes related to pathological findings are underrepresented. Therefore, we conduct experiments on the recognition of such critical RBC subtypes by training linear probing and K-nearest neighbors (K-NN) classifiers on the extracted features. The linear probing assesses the adaptability of our features to downstream tasks, while the K-NN evaluates the effectiveness and robustness of the features under potential batch effect. 

For evaluation, we employ accuracy (Acc), balanced accuracy (bAcc), and the weighted F1 measure (wF1) as metrics.

\begin{table}[ht]
\centering
\caption{RedDino models outperform ResNet50, DINOv2, and DinoBloom by over $2.1\%$ in linear probing evaluation and $3.0\%$ in 1-NN and 20-NN evaluation on the Elsafty dataset using a leave-one-source-out strategy, where one source is fixed for training and the others have been used for testing.}
\resizebox{0.92\textwidth}{!}{
\begin{tabular}{l|ccc|ccc|ccc}
\hline
\textbf{Model} & \multicolumn{3}{c|}{\textbf{Linear probing}} & \multicolumn{3}{c|}{\textbf{1-NN}} & \multicolumn{3}{c}{\textbf{20-NN}} \\ \hline
 & \textbf{wF1} & \textbf{bAcc} & \textbf{Acc} & \textbf{wF1} & \textbf{bAcc} & \textbf{Acc} & \textbf{wF1} & \textbf{bAcc} & \textbf{Acc} \\ \hline
ResNet50~\cite{he_deep_2016} & 77.6{\scriptsize$\pm$8.1} & 80.3{\scriptsize$\pm$6.0} & 77.8{\scriptsize$\pm$8.0} & 64.3{\scriptsize$\pm$4.8} & 65.8{\scriptsize$\pm$4.7} & 64.2{\scriptsize$\pm$4.9} & 66.2{\scriptsize$\pm$4.9} & 66.9{\scriptsize$\pm$4.9} & 67.5{\scriptsize$\pm$4.5} \\ 
DinoBloom-S & 83.2{\scriptsize$\pm$8.2} & 85.2{\scriptsize$\pm$6.7} & 83.3{\scriptsize$\pm$8.0} & 73.1{\scriptsize$\pm$5.1} & 76.7{\scriptsize$\pm$3.9} & 73.2{\scriptsize$\pm$5.1} & 76.5{\scriptsize$\pm$4.2} & 79.6{\scriptsize$\pm$3.6} & 77.1{\scriptsize$\pm$4.1} \\ 
DinoBloom-B & 84.6{\scriptsize$\pm$6.5} & 85.4{\scriptsize$\pm$6.3} & 84.7{\scriptsize$\pm$6.5} & 72.4{\scriptsize$\pm$6.2} & 75.6{\scriptsize$\pm$5.4} & 72.2{\scriptsize$\pm$6.3} & 76.1{\scriptsize$\pm$6.1} & 78.9{\scriptsize$\pm$5.6} & 76.8{\scriptsize$\pm$5.8} \\ 
DinoBloom-L & 85.4{\scriptsize$\pm$5.2} & 87.2{\scriptsize$\pm$4.0} & 85.6{\scriptsize$\pm$5.0} & 74.1{\scriptsize$\pm$5.0} & 76.7{\scriptsize$\pm$4.0} & 74.2{\scriptsize$\pm$4.8} & 77.0{\scriptsize$\pm$4.5} & 79.0{\scriptsize$\pm$4.4} & 77.9{\scriptsize$\pm$4.1} \\ 
DINOv2 small\ & 82.1{\scriptsize$\pm$8.2} & 83.9{\scriptsize$\pm$6.7} & 82.2{\scriptsize$\pm$8.2} & 73.5{\scriptsize$\pm$4.8} & 75.6{\scriptsize$\pm$4.3} & 73.5{\scriptsize$\pm$4.7} & 77.2{\scriptsize$\pm$4.6} & 78.5{\scriptsize$\pm$4.5} & 77.9{\scriptsize$\pm$4.5} \\
DINOv2 base\ & 85.4{\scriptsize$\pm$5.5} & 86.8{\scriptsize$\pm$4.5} & 85.4{\scriptsize$\pm$5.6} & 75.5{\scriptsize$\pm$4.3} & 76.2{\scriptsize$\pm$4.8} & 75.6{\scriptsize$\pm$4.2} & 79.2{\scriptsize$\pm$4.8} & 78.6{\scriptsize$\pm$5.9} & 79.9{\scriptsize$\pm$4.3} \\ 
DINOv2 large\ & 86.0{\scriptsize$\pm$5.6} & 87.2{\scriptsize$\pm$4.6} & 85.9{\scriptsize$\pm$5.7} & 73.7{\scriptsize$\pm$6.2} & 73.3{\scriptsize$\pm$6.5} & 73.9{\scriptsize$\pm$6.2} & 76.4{\scriptsize$\pm$7.0} & 74.8{\scriptsize$\pm$7.3} & 77.4{\scriptsize$\pm$6.4} \\ \hline
RedDino small & 86.0{\scriptsize$\pm$7.0} & 87.2{\scriptsize$\pm$6.1} & 86.2{\scriptsize$\pm$6.6} & 76.8{\scriptsize$\pm$4.9} & \underline{79.8{\scriptsize$\pm$3.4}} & 76.9{\scriptsize$\pm$4.8} & 80.0{\scriptsize$\pm$4.5} & \underline{82.6{\scriptsize$\pm$3.4}} & 80.4{\scriptsize$\pm$4.4} \\ 
RedDino base & \underline{88.1{\scriptsize$\pm$4.9}} & \textbf{89.3{\scriptsize$\pm$4.7}} & \underline{88.2{\scriptsize$\pm$4.9}} & \textbf{78.8{\scriptsize$\pm$3.6}} & \textbf{81.8{\scriptsize$\pm$2.7}} & \textbf{78.6{\scriptsize$\pm$3.7}} & \textbf{82.6{\scriptsize$\pm$2.8}} & \textbf{84.8{\scriptsize$\pm$2.5}} & \textbf{82.9{\scriptsize$\pm$2.8}} \\ 
RedDino large & \textbf{88.5{\scriptsize$\pm$5.5}} & \underline{89.1{\scriptsize$\pm$5.2}} & \textbf{88.4{\scriptsize$\pm$5.5}} & \underline{78.5{\scriptsize$\pm$4.6}} & 79.7{\scriptsize$\pm$4.5} & \underline{78.4{\scriptsize$\pm$4.6}} & \underline{81.6{\scriptsize$\pm$4.7}} & 81.9{\scriptsize$\pm$4.9} & \underline{81.9{\scriptsize$\pm$4.6}} \\ \hline
\textbf{Improvement} & 2.5 & 2.1 & 2.6 & 3.2 & 5.0 & 3.1 & 3.4 & 5.2 & 3.0 \\ 
\textbf{Avg improvement} & 4.0 & 3.4 & 4.1 & 5.6 & 6.1 & 5.6 & 5.9 & 6.5 & 5.3 \\ \hline
\end{tabular}
}
\label{tab:elsafty}
\end{table}

\textbf{Elsafty evaluation} is designed in a cross-source fashion since the Elasfty dataset is subdivided into four sources. This cross-source approach is particularly relevant in clinical trials, where significant variability is often observed in day-to-day analyses due to differences in equipment, protocols, or sample preparation methods. Our strategy involves training shallow models on the features from a fixed source and iteratively testing on the remaining three, cycling through all possible combinations. This choice ensures a comprehensive evaluation of the model's robustness and adaptability to such variations. The results are averaged across the experiments, and the findings are reported in~\Cref{tab:elsafty}. 
In the summary table, we present both the results and the improvements achieved by our RedDino models compared to the selected baselines. Specifically, on the one hand, the row \textit{Improvement} represents the difference between the best results from our RedDino models and the best baseline result. On the other hand, the row \textit{Avg Improvement} captures the difference between the average performance of our models and the average performance of the baselines. 
\begin{table}[ht]
\centering
\caption{RedDino outperforms baseline models in linear probing evaluations, with the only exception of the bAcc on the DSE dataset, in a five-fold cross-validation approach.}
\resizebox{0.63\textwidth}{!}{
\begin{tabular}{l|ccc|ccc}
\hline
\textbf{Name} & \multicolumn{3}{c|}{\textbf{Linear probing Chula}} & \multicolumn{3}{c}{\textbf{Linear probing DSE}} \\ 
\hline
 & \textbf{wF1} & \textbf{bAcc} & \textbf{Acc} & \textbf{wF1} & \textbf{bAcc} & \textbf{Acc} \\ \hline
ResNet50~\cite{he_deep_2016} & 76.9{\scriptsize$\pm$0.4} & 70.7{\scriptsize$\pm$0.6} & 76.9{\scriptsize$\pm$0.4} & 83.2{\scriptsize$\pm$0.8} & 51.5{\scriptsize$\pm$9.0} & 83.2{\scriptsize$\pm$0.7} \\
DinoBloom-S & 81.4{\scriptsize$\pm$0.6} & 76.0{\scriptsize$\pm$1.0} & 81.4{\scriptsize$\pm$0.6} & 84.4{\scriptsize$\pm$1.2} & 57.5{\scriptsize$\pm$3.7} & 84.3{\scriptsize$\pm$1.1} \\
DinoBloom-B & 79.0{\scriptsize$\pm$0.4} & 74.3{\scriptsize$\pm$0.5} & 78.9{\scriptsize$\pm$0.4} & 85.3{\scriptsize$\pm$1.0} & 59.9{\scriptsize$\pm$5.4} & 85.4{\scriptsize$\pm$1.0} \\
DinoBloom-L & 80.0{\scriptsize$\pm$0.3} & 75.0{\scriptsize$\pm$0.7} & 80.2{\scriptsize$\pm$0.6} & \underline{86.2{\scriptsize$\pm$0.8}} & \textbf{60.7{\scriptsize$\pm$3.6}} & \underline{86.2{\scriptsize$\pm$0.9}} \\
DINOv2 small & 81.0{\scriptsize$\pm$0.4} & 73.9{\scriptsize$\pm$0.7} & 81.1{\scriptsize$\pm$0.4} & 83.6{\scriptsize$\pm$0.9} & 54.4{\scriptsize$\pm$3.6} & 83.6{\scriptsize$\pm$0.9} \\
DINOv2 base & 80.1{\scriptsize$\pm$0.5} & 73.6{\scriptsize$\pm$0.5} & 80.1{\scriptsize$\pm$0.6} & 84.8{\scriptsize$\pm$0.5} & 56.2{\scriptsize$\pm$4.5} & 84.8{\scriptsize$\pm$0.5} \\
DINOv2 large & 81.5{\scriptsize$\pm$0.7} & 74.8{\scriptsize$\pm$0.8} & 81.3{\scriptsize$\pm$0.2} & 84.7{\scriptsize$\pm$0.6} & 54.4{\scriptsize$\pm$3.9} & 84.9{\scriptsize$\pm$0.5} \\
\hline
RedDino small & \textbf{84.3{\scriptsize$\pm$0.4}} & \underline{78.5{\scriptsize$\pm$1.1}} & \underline{84.4{\scriptsize$\pm$0.4}} & 84.9{\scriptsize$\pm$1.0} & 56.5{\scriptsize$\pm$5.1} & 84.9{\scriptsize$\pm$0.8} \\
RedDino base & 83.8{\scriptsize$\pm$0.5} & 78.6{\scriptsize$\pm$1.0} & 83.8{\scriptsize$\pm$0.5} & 85.9{\scriptsize$\pm$0.5} & 57.9{\scriptsize$\pm$3.0} & 86.0{\scriptsize$\pm$0.5} \\
RedDino large & \underline{83.9{\scriptsize$\pm$0.5}} & 
\textbf{79.0{\scriptsize$\pm$0.8}} & \textbf{85.0{\scriptsize$\pm$0.4}} & \textbf{86.6{\scriptsize$\pm$1.0}} & \underline{60.1{\scriptsize$\pm$4.8}} & \textbf{86.6{\scriptsize$\pm$1.0}} \\
\hline
\textbf{Improvement} & 2.9 & 3.0 & 3.6 & 0.4 & -0.6 & 0.4 \\ 
\textbf{Avg improvement} & 4.0 & 3.0 & 4.4 & 1.2 & 1.8 & 1.2 \\
\hline
\end{tabular}
}
\end{table}

The \textit{Improvement} metric provides a more impactful depiction of the advancement over the state-of-the-art, while the \textit{Avg Improvement} offers a subtler analysis. Often, the choice of a model for downstream tasks is guided by extensive and computationally expensive benchmarks designed to identify the optimal model through a brute-force search approach. In contrast, the proposed indicator offers a more realistic reflection of performance improvement. Our models demonstrated an improvement of over $2\%$ in reference metrics for the linear probing and over $3\%$ for the 1-NN and 20-NN, surpassing the current state-of-the-art methods. 

\textbf{Chula and DSE evaluation} is designed to represent heavily unbalanced scenarios. We employ a 5-fold cross-validation for both datasets. For the sake of this evaluation, it is worth noting that the Chula dataset was part of the training data for the DinoBloom methods. Despite this, RedDino proves its generalization capabilities, surpassing all other models in linear probing. We do not report results for NN classification due to insufficient samples of rare classes to perform significant analysis. These strong capabilities are also evident in the DSE dataset, where RedDino outperforms all other models in almost all the selected metrics and evaluation algorithms. 
\newline\newline
It is important to note that for all analyzed datasets, the \textit{Avg Improvement} was always positive, showcasing the importance and relevance of our models, as they consistently outperformed other representations for RBC analysis tasks across different microscopes and sources.
Additionally, the evaluations yielded key insights: training RedDino on patched smear images rather than individual cells significantly enhanced model performance. Furthermore, RedDino base proved to be a strong general solution, balancing performance with efficiency by utilizing fewer parameters (86 million vs. 304 million).

\begin{figure}[htb]
    \centering
    \caption{Abnormal RBC distinguished by RedDino: the highlighted regions in (a) and (c) correlate with distinct colors in the PCA visualization (b) and (d), showcasing their differentiation provided in the embedding space. Specifically, (a) contains malaria-infected RBCs, while (c) includes echinocytes.}
    \includegraphics[width=0.45\linewidth]{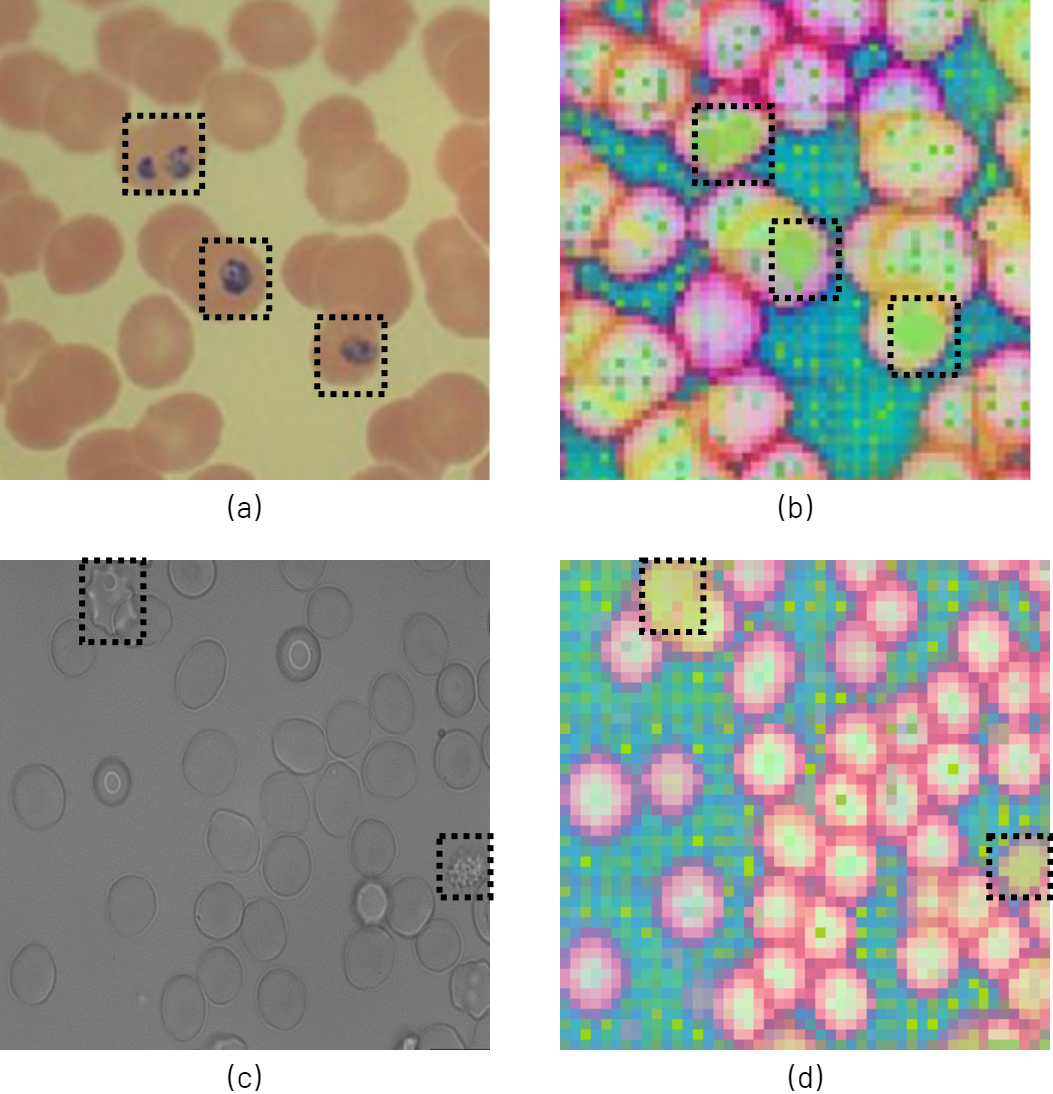}    
    \label{fig:pca}
\end{figure}

\subsection{Visualization of the RedDino Features}
\textbf{PCA Visualization} demonstrates feature relevance via two analyses of RedDino features in a three-components PCA visualization (\Cref{fig:pca}). 
Panels (a)-(b) show a smear patch from the MP-IDB dataset with healthy and \textit{Plasmodium falciparum}-infected RBCs and the PCA visualization from which it is observable how the model distinguishes background, cells, membranes, and parasites with high responses, highlighting the latter structures. 
Panels (c)-(d) apply the same approach to a brightfield image, revealing distinct features for echinocytes from the DSA dataset. This behavior arises from self-supervised training alone. The regions of interest are marked with black dotted boxes.

\begin{figure}
    \centering
    \caption{Different classes show distinct clusters in the UMAP projection of the feature embeddings from the Elsafty dataset source 1. On the left, we show the subject distribution across the UMAP space, while on the right, we show the class distribution.}
    \includegraphics[width=1\linewidth]{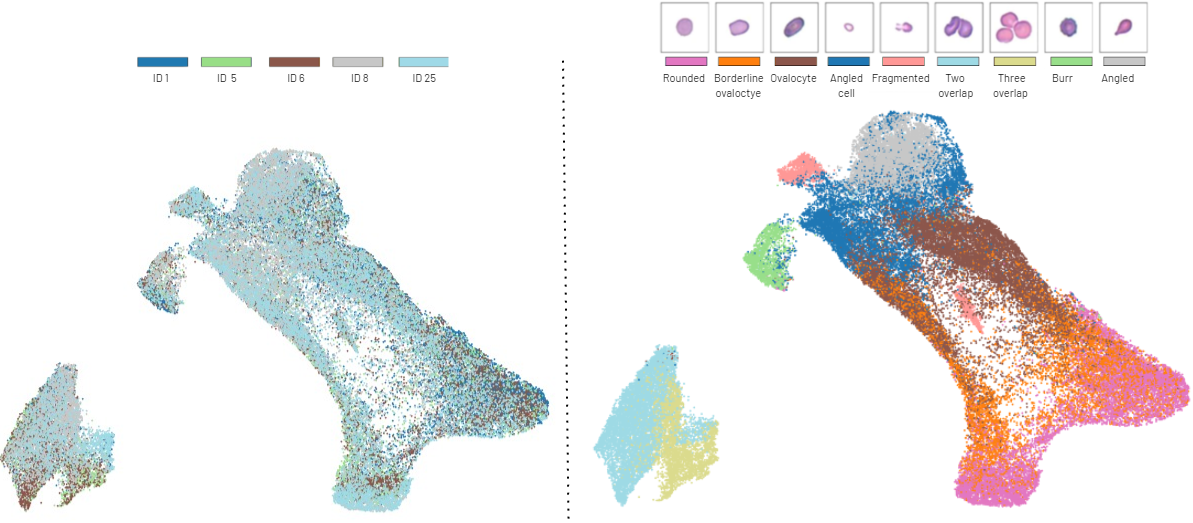}
    
    \label{fig:umap}
\end{figure}

\noindent\textbf{UMAP Visualization} reveals batch effects within the feature space. Using the Elsafty dataset (source 1), \Cref{fig:umap} shows distinct clusters without subclusters, indicating no batch across individuals. Nevertheless, overlapping classes (Rounded RBCs, Ovalocytes, and Borderline Ovalocytes) remain hard to separate due to the lack of clear clinical thresholds. Clumps form distinct clusters, confirming the model’s ability to differentiate single cells from agglomerations.
\newline\newline
\noindent\textbf{CO2 Emissions from Experiments.}
Our experiments were run on our in-house infrastructure, using the NVIDIA A100SXM4-80GB hardware; the total emissions are estimated to be 4.15 kg CO2eq.

\section{Conclusion}
In this paper, we presented RedDino, a cutting-edge foundation model specifically designed for red blood cell analysis. By leveraging self-supervised learning techniques, particularly a custom DINOv2 architecture, our models demonstrated superior generalization across a wide range of red blood cell datasets, effectively mitigating challenges such as the batch effect. The extensive range of experimentations conducted, including an extensive comparative investigation and multiple scenario evaluations, confirm that RedDino achieves state-of-the-art performance in red blood cell classification, surpassing previous baselines. The model's ability to adapt to different data sources, imaging protocols, and patient populations demonstrates its robustness and potential for clinical applications. RedDino sets a new standard for RBC analysis and offers a strong foundation for future advancements in automated hematological diagnostics.

%
%
%

\bibliographystyle{splncs04}  
\bibliography{references}

\end{document}